\documentclass[10pt,twocolumn,letterpaper]{article}
\usepackage{times}
\usepackage{epsfig}
\usepackage{graphicx}
\usepackage{amsmath}
\usepackage{amssymb}
\usepackage{color}
\usepackage[dvipsnames]{xcolor}
\usepackage{multirow}
\usepackage{bbm}
\usepackage{booktabs}
\usepackage{graphicx}
\usepackage{placeins}
\usepackage[toc]{appendix}
\usepackage{mathtools}

\graphicspath{{img/}}
\date{}
\usepackage[pagebackref=true,breaklinks=true,letterpaper=true,colorlinks,bookmarks=false]{hyperref}

\newcommand{\N}[0]{\mathbb{N}}
\newcommand{\Z}[0]{\mathbb{Z}}
\renewcommand{\leq}{\leqslant}

\usepackage{adjustbox}

\renewcommand{\baselinestretch}{1.0}

\begin{document}

\title{Closed-form Quadrangulation of $n$-Sided Patches} 

\author{
\parbox{1.0\textwidth}{
\centering
Marco Tarini \\
Università degli Studi di Milano\\
{\tt\small marco.tarini@unimi.it}}
}

\maketitle

\begin{abstract}
We analyze the problem of quadrangulating a $n$-sided patch, each side at its boundary subdivided into a given number of edges, using a single irregular vertex (or none, when $n = 4$) that breaks the otherwise fully regular lattice. We derive, in an analytical closed-form, (1) the necessary and sufficient conditions that a patch must meet to admit this quadrangulation, and (2) a full description of the resulting tessellation(s).
\end{abstract}

\section{Introduction}

Consider a polygonal-shaped, planar region patch $P$, delimited by $n>1$ sides, each subdivided in a number of edges. Let $e_i \in \N$ be the number of edges found on side $i$, $i \in [0..n-1]$.

We are interested in determining whether or not $P$ can be quad-tessellated using only one irregular vertex, of valency $n$, somewhere in the interior (even this vertex is regular when $n=4$). 

This tessellation, when it exists, can also be described as the one obtained 
by applying one step of Catmull-Clark (CC) subdivision \cite{CC} to the polygon $P$, which creates $n$ quadrilateral regions, followed by a conforming, fully regular tessellation of each of these regions, each at some appropriate grid resolution.
For this reason, we refer to this property of $P$ as it being ``CC-able''.

\begin{figure}
    \centering
    \includegraphics[width=0.9\columnwidth]{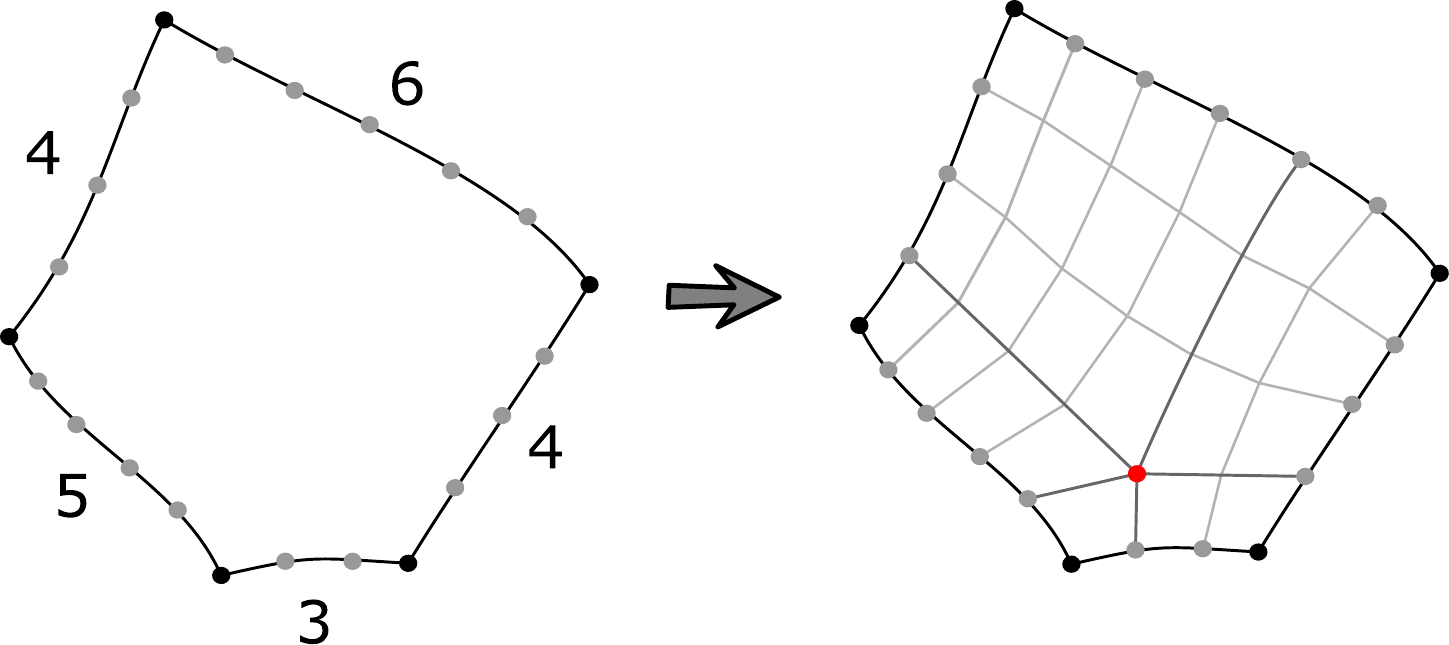}
    \caption{An example: a pentagonal patch with $e_{0..4} = (6,4,3,5,4)$ is CC-able, as graphically  shown here.}
    \label{fig:example}
\end{figure}

For example, the construction in Fig. \ref{fig:example} shows that a pentagonal-shaped patch delimited by $e_i = (6,4,3,5,4)$ edges happens to be CC-able.
We show a closed-form formulation to determine the existence (and the uniqueness) of this quadrangulation, for an arbitrary $P$, and to construct it, when it exists.
%Would a patch with with $(8,6,4,3,5)$ be CC-able?

\subsection{Motivations and context}
\label{sec:motivations}

An intensively studied recent problem in Geometry Processing is \textbf{Coarse layout decomposition}
(see \cite{Campen2017a} for a survey), where a given surface must be partitioned into a layout of patches
homologous to a disk. This is a precondition for numerous tasks (such as surface parametrization, shape segmentation, shape matching, application of machine learning on 3D shapes), each
coming with its own set of useful final applications (ranging from texture mapping, digital fabrication, shape recognition, shape modelling, etc). Clearly, each such scenario dictates different requirements on the layout. Among others, one central application is surface semi-regular quad-remeshing (see \cite{Bommes2013} for a survey, and specifically Section 3.2 there).
For other potential contexts, %requiring a similar setup, 
see the Conclusions. 

The layout is typically required to be comprised of rectangular regions, as they admit a ``natural'' tessellation consisting of a fully-regular lattice -- but only if opposite sides of the region are subdivided in an equal number of edges.

% This observation is at the basis of 
%

The present work can be understood as a way to \emph{extend the concept of a ``natural tessellation'' to non-rectangular patches}. 

\begin{figure*}
    \centering
\includegraphics[width=1.0\textwidth]{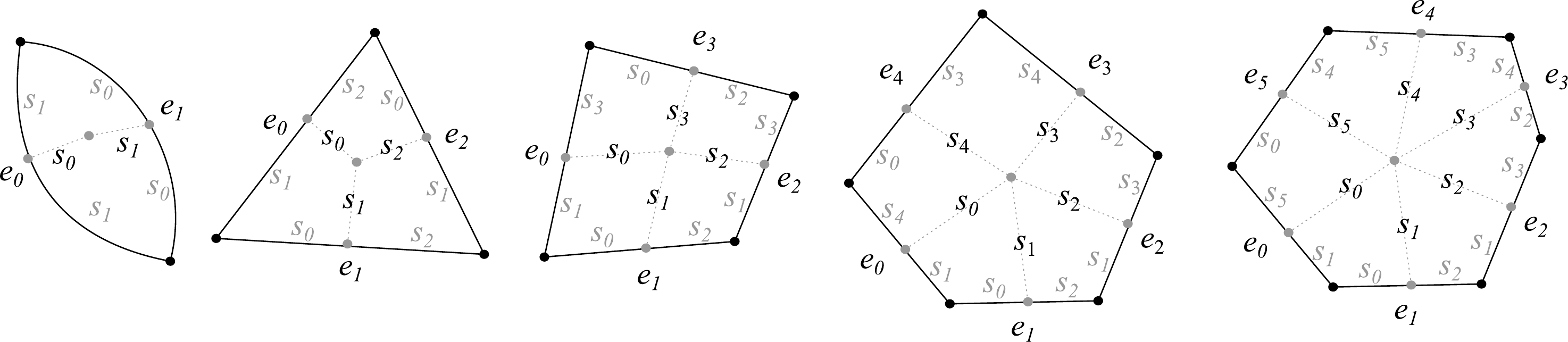}
    \caption{Schema for parameters $e_i$ and unknowns $s_i$ for patches with $n$ = 2, 3, 4, 5 and 6 sides. The same schema generalises to larger $n$.}
    \label{fig:schema}
\end{figure*}

\section{Related work}

The literature addresses a problem similar to ours, but where \textbf{multiple} internal irregular vertices are allowed. In this case, solutions are now known to always exists  \cite{Takayama2014} (as long as the ``parity'' condition holds, see below); note that layouts presenting more internal irregular vertices are typically considered less desirable in many contexts. 

This version of the problem was solved in \cite{Takayama2014} for $n \leq 6$, and an algorithm is offered that always constructs one valid solution. The problem was previously tackled  for $n = 3$ or $4$, with similar results \cite{takayama2013robust}. These algorithms are combinatorial in nature; this is in contrast with our closed-form characterization of the instances admitting a solution with a single-irregular vertex (in addition to solving also for $n>6$).

The algorithm in \cite{Takayama2014} returns the one solution considered the ``best'' among valid alternatives, according to some definition of desirability (although not all possibilities are always considered). In a similar spirit, 
Machine-Learning  \cite{matveev2020def} and Data-Driven \cite{Marcias2015} heuristics have been leveraged to pick one ``best'' solution among the ones admitted by a given instance (again, according to some targeted criteria). In older proposals, users, such as digital modellers, are allowed to interactively navigate inside the space of admissible solutions, in search of the ``preferred'' one for a given mesh \cite{editMesh2011}. In most cases (although not necessarily all), the ``best'' or ``preferred'' solution is exactly the one featuring a single irregular vertex inside the patch, when such a solution exists; to our knowledge, our is the first criterion to determine \emph{a priori} whether that is the case or not.

\section{Solution}

One well-known precondition for $P$ to be tasselable with only quads is that %
\begin{equation}
\left( \sum_i{e_i} \right) \text{mod}\, 2 = 0.
\label{eq:even}
\end{equation}
We refer to this as the \emph{parity condition}, and we assume it always holds.

We are looking for a tessellation with a single, internal irregular vertex.
In the only possible construction for a solution is depicted in  Figure \ref{fig:schema}, where each side around $P$ is split into two sub-sides. 
Due to the constraint on regularity, a pair of sub-sides at the left and the right of another edge $i$ must share the same number of edges $s_i$.

Thus, we have that
\begin{equation}
\forall i < n, \; \;  e_i = s_{i-1} + s_{i+1}
\label{eq:edgesbasic}
\end{equation}
(all indices, in the above and in all following equations, are considered modulo $n$).

Because each side of the polygon must be split in two sub-sides, we also need to assume $e_i>1$.

For $P$ to be CC-able, equations \ref{eq:edgesbasic} must be fulfilled for some choice of unknown positive integer values $s_i$. This general property translates into different sets of conditions for each value of $n$, as we analyze in the following sections. 

\subsection{Two-sided shapes ($n=2$)}
For $n=2$, the problem is trivial. Equation (\ref{eq:edgesbasic}) becomes simply
\begin{equation}
e_0 = 2 \; s_{1}  \;\;\;\; \text{ , }  \;\;\;\;  e_1 = 2 \; s_{0}
\label{eq:twoSi}
\end{equation}
(as $i-1$ and $i+1$ denote the same index modulo 2).  

\begin{figure}
\centering
\includegraphics[width=0.25\columnwidth]{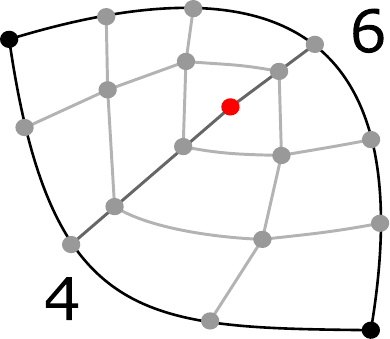}
    \caption{A CC-able 2-sided patch, shown with its the corresponding internal quadrangulation.}
    \label{fig:example2}
\end{figure}

Note that it is not necessarily the case that valency-2 vertices (sometimes called \emph{doublets}, \cite{QuadMeshSimp}) are considered valid configurations, but they can be \cite{QuadMeshSimp2}. See also Fig. \ref{fig:example2}.
%on a plane (but not in presence of curvature), a doublet implies either two quads with flat angles, or a concave quad.

In conclusion, \emph{a two-sided patch is CC-able (in only one way) iff both $e_0$ and $e_1$ are even numbers} (a stronger condition than the Parity condition).

\subsection{Triangular shapes ($n=3$)}

For $n=3$, we rewrite Equation (\ref{eq:edgesbasic}) in matrix form: 
\begin{equation}
\left( \begin{matrix} 
 0 & 1 & 1  \\
 1 & 0 & 1  \\ 
 1 & 1 & 0  
 \end{matrix} \right)
\left( \begin{matrix} 
 s_0 \\
 s_1 \\ 
 s_2  
\end{matrix} \right)
 =
\left( \begin{matrix} 
 e_0 \\
 e_1 \\ 
 e_2 
\end{matrix} \right)
\end{equation}
 which implies (by matrix inversion):
\begin{equation}%
 \frac{1}{2} 
\left( \begin{matrix} 
 -1 & +1 & +1 \\
 +1 & -1 & +1 \\ 
 +1 & +1 & -1
\end{matrix} \right)
\left( \begin{matrix} 
 e_0 \\
 e_1 \\ 
 e_2 
\end{matrix} \right)
 =
\left( \begin{matrix} 
 s_0 \\
 s_1 \\ 
 s_2 
\end{matrix} \right)
\end{equation}

In other terms, 
\begin{equation}
\forall i , \; \;  s_i = \frac{1}{2} \Big(  e_{i-1} - e_{i} + e_{i+1}   \Big). 
\label{eq:triSi}
\end{equation}

For $s_i$ to be integer, the expression in parenthesis must be even, which is already guaranteed by Eq. (\ref{eq:even}).

%-- the same consideration applies to all other cases below too.

For $s_i$ to be positive, we also need the above expression to be positive, therefore:
\begin{equation}
\forall i , \; \;  e_{i-1} + e_{i+1} >  e_{i}. 
\label{eq:triDis}
\end{equation}

In conclusion, \emph{a triangular patch is CC-able iff each side has fewer edges than the other two sides combined.}
This condition is a discrete version of the familiar triangle inequality. 

%In conclusion, \emph{a triangular patch is CC-able iff its sides fulfill the strict triangle inequality (i.e., each side has fewer edges than the other two combined).}

\subsection{Quadrilateral shapes ($n=4$)}

For a quadrilateral shape, to be CC-able amounts to be regularly griddable. The conditions for this to be the case are well-known and obvious, and are re-derived here (consistently with the other cases) only for completeness.

With $n=4$, Eq. (\ref{eq:edgesbasic}) can be written as
\begin{equation}
\left( \begin{matrix} 
 0 & 1 & 0 & 1 \\ 
 1 & 0 & 1 & 0 \\
 0 & 1 & 0 & 1 \\
 1 & 0 & 1 & 0  
\end{matrix} \right)
\left( \begin{matrix} 
 s_0 \\
 s_1 \\ 
 s_2 \\ 
 s_3 
\end{matrix} \right)
 =
\left( \begin{matrix} 
 e_0 \\
 e_1 \\ 
 e_2 \\ 
 e_3 
\end{matrix} \right)
\end{equation}

The matrix is non-singular, as the first two rows match the second two; therefore, the system has either multiple solutions for $s_i$, when $e_0 = e_2$ and $e_1 = e_3$, or no solution otherwise. This condition also implies the parity condition, as $\sum e_i = 2(e_0+e_1)$.

Because $n=4$, the internal ``irregular'' vertex is, actually, regular like the others, and every valid choice of $s_i$ produces the same tessellation.

In conclusion, \emph{a quadrilateral patch is CC-able (in only one way) when the opposite sides have a matching number of edges.}

\subsection{Pentagonal shapes ($n=5$)}

For $n=5$, we rewrite Eq. (\ref{eq:edgesbasic}) as
\begin{equation}
\left( \begin{matrix} 
 0 & 1 & 0 & 0 & 1 \\
 1 & 0 & 1 & 0 & 0 \\ 
 0 & 1 & 0 & 1 & 0 \\ 
 0 & 0 & 1 & 0 & 1 \\ 
 1 & 0 & 0 & 1 & 0 
\end{matrix} \right)
\left( \begin{matrix} 
 s_0 \\
 s_1 \\ 
 s_2 \\ 
 s_3 \\ 
 s_4 
\end{matrix} \right)
 =
\left( \begin{matrix} 
 e_0 \\
 e_1 \\ 
 e_2 \\ 
 e_3 \\ 
 e_4 
\end{matrix} \right)
\end{equation}
 which implies (by matrix inversion)
\begin{equation}%
 \frac{1}{2} 
\left( \begin{matrix} 
 +1 & +1 & -1 & -1 & +1 \\
 +1 & +1 & +1 & -1 & -1 \\ 
 -1 & +1 & +1 & +1 & -1 \\ 
 -1 & -1 & +1 & +1 & +1 \\ 
 +1 & -1 & -1 & +1 & +1 
\end{matrix} \right)
\left( \begin{matrix} 
 e_0 \\
 e_1 \\ 
 e_2 \\ 
 e_3 \\ 
 e_4 
\end{matrix} \right)
 =
\left( \begin{matrix} 
 s_0 \\
 s_1 \\ 
 s_2 \\ 
 s_3 \\ 
 s_4 
\end{matrix} \right)
\end{equation}

In other terms, 
\begin{equation}
\forall i , \; \;  s_i = \frac{1}{2} \Big( \left( e_{i-1} + e_{i} + e_{i+1} \right) - \left( e_{i-2} + e_{i+2} \right)  \Big).
\label{eq:pentaSi}
\end{equation}

For $s_i$ to be integer, the expressions above must be even before halving, which is already guaranteed by Eq. (\ref{eq:even}).

The sign constraint ($s_i>0$) produces
\begin{equation}
\forall i , \; \;  e_{i-1} + e_{i} + e_{i+1} >  e_{i-2} + e_{i+2} 
\label{eq:pentaDis}
\end{equation}

In conclusion, \emph{a pentagonal patch is CC-able (in only one way) iff the total number of edges in any three consecutive edges is larger than the number of edges in the other two.}
 %
 % The solution, if it exists, is unique.

\subsection{Hexagonal shapes ($n=6$)}

For $n=6$, we rewrite Eq. (\ref{eq:edgesbasic}) as
\begin{equation}
\left( \begin{matrix} 
 0 & 1 & 0 & 0 & 0 & 1 \\
 1 & 0 & 1 & 0 & 0 & 0 \\ 
 0 & 1 & 0 & 1 & 0 & 0 \\ 
 0 & 0 & 1 & 0 & 1 & 0 \\ 
 0 & 0 & 0 & 1 & 0 & 1 \\
 1 & 0 & 0 & 0 & 1 & 0 
\end{matrix} \right)
\left( \begin{matrix} 
 s_0 \\
 s_1 \\ 
 s_2 \\ 
 s_3 \\ 
 s_4 \\ 
 s_5 
\end{matrix} \right)
 =
\left( \begin{matrix} 
 e_0 \\
 e_1 \\ 
 e_2 \\ 
 e_3 \\ 
 e_4 \\ 
 e_5 
\end{matrix} \right)
\end{equation}
 which implies (by matrix inversion)
\begin{equation}%
 \frac{1}{2} 
\left( \begin{matrix*}[r]
   0 & +1 &  0  & -1 &  0 & +1 \\
  +1 &  0 & +1  &  0 & -1 &  0 \\ 
   0 & +1 &  0  & +1 &  0 & -1 \\ 
  -1 &  0 & +1  &  0 & +1  & 0  \\ 
   0 & -1 &  0  & +1 &  0 & +1 \\ 
  +1 &  0 & -1  &  0 & +1 &  0 \\ 
\end{matrix*} \right)
\left( \begin{matrix} 
 e_0 \\
 e_1 \\ 
 e_2 \\ 
 e_3 \\ 
 e_4 \\ 
 e_5 
\end{matrix} \right)
 =
\left( \begin{matrix} 
 s_0 \\
 s_1 \\ 
 s_2 \\ 
 s_3 \\ 
 s_4 \\ 
 s_5  
\end{matrix} \right)
\end{equation}

In other terms, 
\begin{equation}
\forall i , \; \;  s_i = \frac{1}{2} \Big(  e_{i-1} + e_{i+1}  -  e_{i+3}  \Big).
\label{eq:hexaSi}
\end{equation}

 %the expressions above must be integer, which
 
The integrity constraint ($s_i \in \Z$)
translates in the requirement
for the sum of $e_{0,2,4}$, and the sum of $e_{1,3,5}$, to be even numbers.

The sign constraint ($s_i>0$) gives, by $j=(i+3)$ modulo 6:
\begin{equation}
\forall j , \; \;  e_{j-2} + e_{j+2} >  e_{j}.
\label{eq:disHexa}
\end{equation}

In conclusion, \emph{a hexagonal patch is CC-able (in only one way) iff  (1)
%the both its three even and its three odd sides fulfill the same criterion for triangles (including parity). 
each even side has fewer edges than the other two even sides combined (and likewise, for odd sides), and (2), both the even sides, and the odd sides, have an even total number of edges.
}

\subsection{Heptagonal shapes ($n=7$)}

For $n=7$, we rewrite Eq. (\ref{eq:edgesbasic}) as
\begin{equation}
\left( \begin{matrix} 
 0 & 1 & 0 & 0 & 0 & 0 & 1 \\
 1 & 0 & 1 & 0 & 0 & 0 & 0 \\ 
 0 & 1 & 0 & 1 & 0 & 0 & 0 \\ 
 0 & 0 & 1 & 0 & 1 & 0 & 0 \\ 
 0 & 0 & 0 & 1 & 0 & 1 & 0 \\
 0 & 0 & 0 & 0 & 1 & 0 & 1\\
 1 & 0 & 0 & 0 & 0 & 1 & 0 
\end{matrix} \right)
\left( \begin{matrix} 
 s_0 \\
 s_1 \\ 
 s_2 \\ 
 s_3 \\ 
 s_4 \\
 s_5 \\ 
 s_6 
\end{matrix} \right)
 =
\left( \begin{matrix} 
 e_0 \\
 e_1 \\ 
 e_2 \\ 
 e_3 \\ 
 e_4 \\ 
 e_5 \\ 
 e_6 
\end{matrix} \right)
\end{equation}
 which implies (by matrix inversion)
 
\small\begin{equation}%
 \frac{1}{2} 
\left( \begin{matrix} 
  -1\!&\!+1\!&\!+1\!&\!-1\!&\!-1\!&\!+1\!&\!+1 \\
  +1\!&\!-1\!&\!+1\!&\!+1\!&\!-1\!&\!-1\!&\!+1 \\
  +1\!&\!+1\!&\!-1\!&\!+1\!&\!+1\!&\!-1\!&\!-1 \\
  -1\!&\!+1\!&\!+1\!&\!-1\!&\!+1\!&\!+1\!&\!-1 \\
  -1\!&\!-1\!&\!+1\!&\!+1\!&\!-1\!&\!+1\!&\!+1 \\
  +1\!&\!-1\!&\!-1\!&\!+1\!&\!+1\!&\!-1\!&\!+1 \\
  +1\!&\!+1\!&\!-1\!&\!-1\!&\!+1\!&\!+1\!&\!-1 
\end{matrix} \right) 
\left( \begin{matrix} 
 e_0 \\
 e_1 \\ 
 e_2 \\ 
 e_3 \\ 
 e_4 \\
 e_5 \\ 
 e_6 
\end{matrix} \right)
 =
\left( \begin{matrix} 
 s_0 \\
 s_1 \\ 
 s_2 \\ 
 s_3 \\ 
 s_4 \\ 
 s_5 \\ 
 s_6  
\end{matrix} \right)
\end{equation}\normalsize

In other terms, 
\begin{align}
\begin{split}
\forall i , \; \;  s_i = \frac{1}{2} \Big( %\left( 
e_{i-2} + e_{i-1} + e_{i+1} + e_{i+2}%\right) 
&\\
- \left(e_{i} + e_{i+3} + e_{i-3} \right) \;\;\;\; &\Big).
\label{eq:eptaSi}
\end{split}
\end{align}
%\end{equation}

For $s_i$ to be integer, the expression in parenthesis must be even, which is already guaranteed by Eq. (\ref{eq:even}).

The sign constraint ($s_i>0$) becomes 
\begin{equation}
\forall i , \; \;  e_{i-2} + e_{i-1} + e_{i+1} + e_{i+2} > e_{i} + e_{i+3} + e_{i-3}.
\label{eq:disHepta}
\end{equation}

In conclusion, \emph{a heptagonal patch is CC-able (in only one way) iff each side plus its two opposite sides have fewer edges than the remaining four sides combined.}
 %
% The solution, if it exists, is unique.

\subsection{Octagonal shapes ($n=8$)}

For $n=8$, we can rewrite Eq. (\ref{eq:edgesbasic}) as two 
separate linear systems:

\begin{equation}
 \begin{split}
\left( \begin{matrix} 
 1 & 0 & 0 & 1  \\
 1 & 1 & 0 & 0  \\ 
 0 & 1 & 1 & 0  \\ 
 0 & 0 & 1 & 1  \\ 
\end{matrix} \right)
\left( \begin{matrix} 
 s_1 \\
 s_3 \\ 
 s_5 \\ 
 s_7 
\end{matrix} \right)
 & =
\left( \begin{matrix} 
 e_0 \\
 e_2 \\ 
 e_4 \\ 
 e_6  
\end{matrix} \right)
 %
%\;\;\text{and}\;\;
\\
\left( \begin{matrix} 
 1 & 1 & 0 & 0  \\
 0 & 1 & 1 & 0  \\ 
 0 & 0 & 1 & 1  \\ 
 1 & 0 & 0 & 1  \\ 
\end{matrix} \right)
\left( \begin{matrix} 
 s_0 \\
 s_2 \\ 
 s_4 \\ 
 s_6 
\end{matrix} \right)
 & =
\left( \begin{matrix} 
 e_1 \\
 e_3 \\ 
 e_5 \\ 
 e_7  
\end{matrix} \right)
 \end{split}
\end{equation}

Neither matrix is invertible, being deficient by one rank: specifically, their two even rows and two odd rows sum up to the same row-vector. Therefore, the system can have solutions only when
\begin{equation}
\begin{split}
 e_0 + e_4 &= e_2 + e_6, \\
 e_1 + e_5 &= e_3 + e_7.
 \end{split}
\label{eq:disOcta}
\end{equation}
In the following, we show that this condition is also sufficient.

The parity condition is already implied by Eq.~(\ref{eq:disOcta}), because $\sum{e_i} = 2 (e_0+e_4+e_1+e_5)$.

The only remaining condition is that $s_i > 0$. Let $k_0, k_1$ be the choices
for $s_0, s_1$ (consented by the rank deficits). 
Using Eq. (\ref{eq:edgesbasic}), we get the values of all other $s_i$:
\begin{align}
%\begin{split}
 s_0 &= k_0,              & s_1 &= k_1 ,         \nonumber
 \\[1.5mm]
 s_2 &= e_1-s_0          & s_3 &= e_2-s_1      \nonumber
  \\[-1mm]  &= e_1-k_0,          &     &= e_2-k_1,      \nonumber
  \\[1.5mm]
 s_4 &= e_3-s_2          & s_5 &= e_4-s_3      \label{eq:octaSi}
  \\[-1mm]  &= e_3-e_1+k_0 ,     &     &= e_4-e_2+k_1,   \nonumber
  \\[1.5mm] 
 s_6 &= e_5-s_4          & s_7 &= e_6-s_5          \nonumber
  \\[-1mm]  &= e_5-e_3+e_1-k_0  &     &= e_6-e_4+e_2-k_1  \nonumber
  \\[-1mm]  &= e_7-k_0 ,         &     &= e_0-k_1 .         \nonumber
% \end{split}
\end{align}
(the bottom line uses Eq. (\ref{eq:disOcta})).
Therefore,  to have $s_i > 0 $, the choices $k_{0}$ and $k_{1}$ must be, each, subject to two upper-bounds and two lower-bounds:
\begin{align}
     0 &< k_0 < e_1,   &  0 &< k_1 < e_2,\;\;\;\;\;\;   \nonumber    \\[-3.mm]
     \label{eq:octaKi}
     \\[-2mm]
     e_1-e_3 &< k_0 < e_7,   &  e_2-e_4 &< k_1 < e_0.\;\;\;\;\;\;   \nonumber    
\end{align}
The above set of constraints is always feasible. The existence of a valid integer solution for $k_0$  (and likewise, for $k_1$)  is guaranteed because each of the two lower bounds is smaller, by at least two units, than each of the two upper bounds. \emph{Proof:} it is immediate to verify this for $0 < e_{1,7}$ and $e_1-e_3<e_1$ (as $e_{1,3,7}>1$); finally, it also holds for $e_1-e_3 < e_7$ because $e_1 < e_7 +e_3 = e_1 + e_5$ (and $e_5>1$; the last equality is  Eq.~\ref{eq:disOcta}).~{\small$ \square\;$}

In conclusion, \emph{an octagonal patch is CC-able (in general, in multiple ways) iff one pair of even opposite sides have the same combined number of edges as the other pair of even opposite sides, and same for odd sides.}

\begin{figure}
\includegraphics[width=1.0\columnwidth]{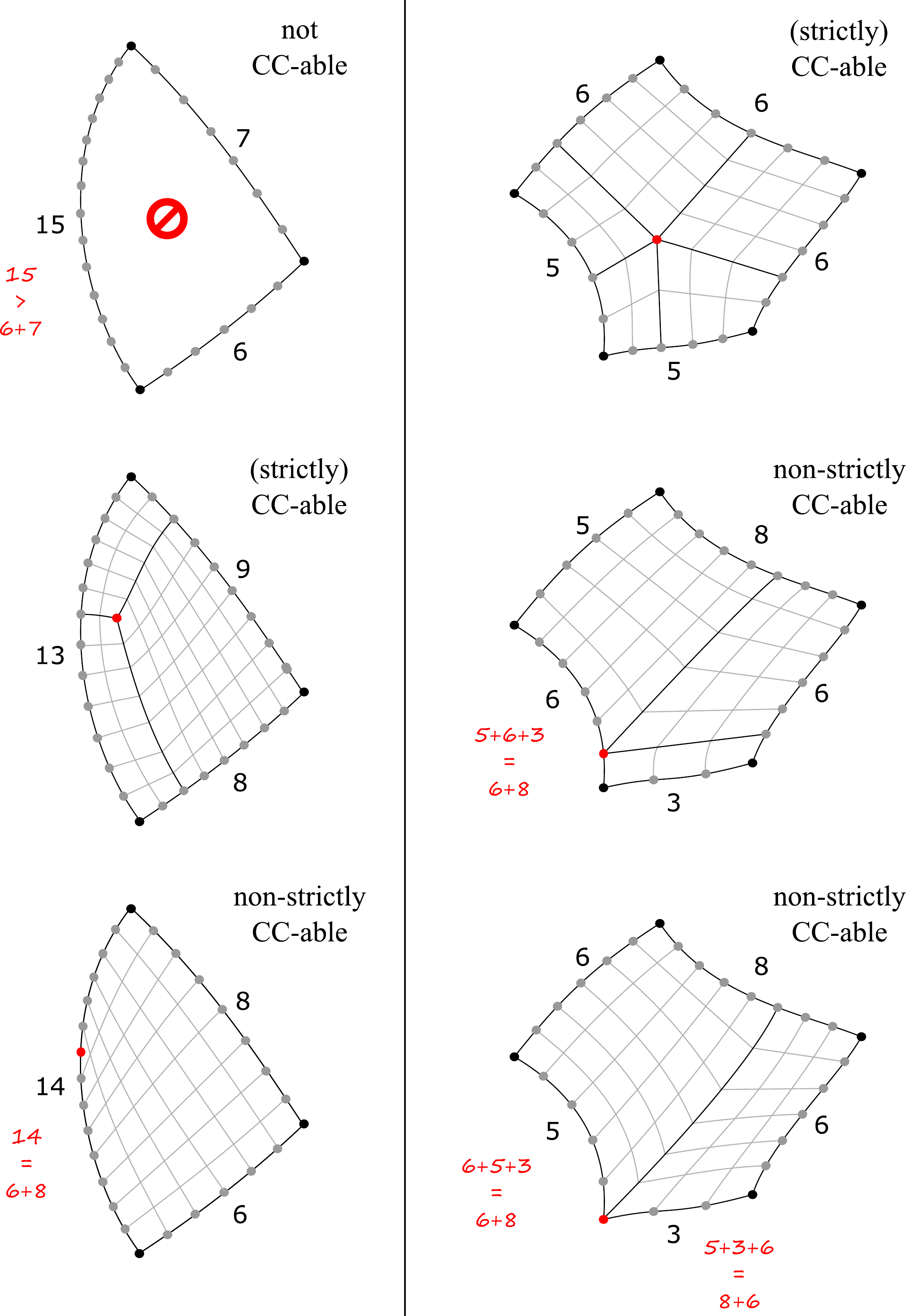}
    \caption{Examples of (strict or non-strict) CC-ability being met or not met for triangular (left) and pentagonal (right) patches. The solitary singular vertex is the red dot; the inequalities (respectively, equations \ref{eq:triDis},\ref{eq:pentaDis}) which are not met, or met in a non-strict sense, are annotated in red. See Section \ref{sec:nonstrict}.
    \label{fig:nonstrict}}
\end{figure}

\subsection{Generalization to $n>8$}

Our CC-ability analysis for each $n \leq 8$
covers (we think) the cases used by most practical scenarios. 

The case for any other $n > 8$  %are straightforward,
can be constructed similarly, presenting only more instances of the situations encountered for $n \leq 8$. We only briefly surmise a generalization here.

For odd values of $n$, Eq.~(\ref{eq:edgesbasic}) can always be expressed with an invertible linear system, 
which can be inverted to extract the values of $s_i$ 
as a linear expression of the form 
\begin{equation}
    s_i = \frac{1}{2}\sum_j \pm e_j > 0
    \label{eq:generalSi}
\end{equation}(as exemplified for $n=3,5,7$).
Because $\sum \pm e_j$ has always the same parity as $\sum e_j$, the condition that $s_i \in \Z$ is already guaranteed by the Parity condition, Eq. (\ref{eq:even}).
The condition for CC-ability is thus entirely determined by the signs of $s_i$, which translates into a set of $n$ linear inequalities, each involving all $e_i$. For a given patch, there is either one or no solution.

For $n = 2h$ ($h \in \N$),  Eq.~(\ref{eq:edgesbasic}) can be expressed as two separate linear systems, partitioning the patch sides in two alternating subsets of $h$ elements each. When $h$ is odd (i.e., $n$ is not a multiple of 4), both systems are invertible, and CC-ability is then determined by a total of $n$ inequalities, each involving $h$ sides, plus the condition that the total number of edges on each subset must be even; the solution is always unique, if it exists.  When $h$ is even (i.e., $n$ is a multiple of 4) neither system is invertible, each being deficient by one rank; the CC-ability is determined by the equality conditions on $e_i$ required by either system to admit solutions. Solutions can be multiple (but equivalent when $n=4$, as noted), and their space is spanned by two degrees of freedom.

\section{Non-strict CC-ability}
\label{sec:nonstrict}
The definition of CC-ability can be relaxed by accepting that the one singularity is found on the boundary of the patch, rather than in its interior (see Figure~\ref{fig:nonstrict} for examples). We term the relaxed definition ``non-strict CC-ability''.

% Which can be acceptable, depending on the context of use. 
For a patch to be non-strictly CC-able, we need $s_i$ to be just non-negative, rather than strictly positive, thus allowing for all the inequalities in the conditions for CC-ability to be fulfilled in the non-strict sense, in Equations (\ref{eq:triDis},\ref{eq:pentaDis},\ref{eq:disHexa},\ref{eq:disHepta},\ref{eq:octaKi}) for the cases $n = $ 3,5,6,7,8, respectively. The assumption that $e_i>1$ can also be dropped, allowing for $e_i = 1$ in the input patch. 

When \emph{one} inequality is fulfilled as equality, the irregular vertex will be on the boundary, i.e.\ as a boundary vertex with edge-valency $\neq 3$.
When \emph{two} inequalities are fulfilled as equalities, the irregular vertex will be found on a corner, i.e. as a corner vertex with edge-valency $\neq 2$ (as shown in bottom-right of Figure~\ref{fig:nonstrict}).
It is never possible for more than two inequalities to be fulfilled as equalities (without infringing at least another inequality).

\begin{figure}
\includegraphics[width=1.0\columnwidth]{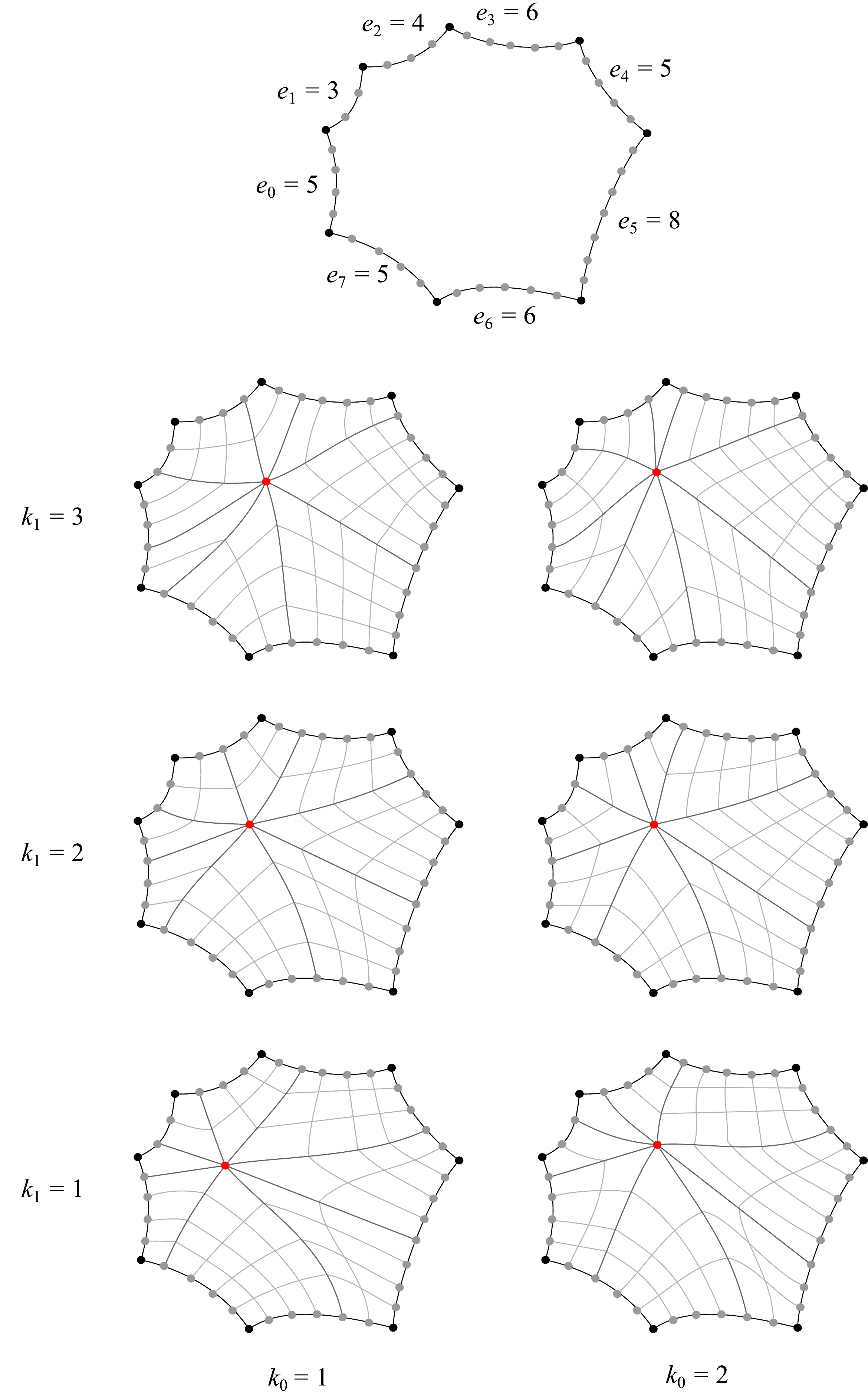}
    \caption{An octagonal patch (top) fulfills the conditions in Eq.~(\ref{eq:disOcta}) and is therefore CC-able. After Eq. (\ref{eq:octaKi}), we have two choices $k_0,k_1$, ranging in the intervals $0 < k_0 < 3$ and $0 < k_1 < 4$; all resulting quadrangulations are shown.}
    \label{fig:example8}
\end{figure}

%We stated all our conditions as strict inequalities, implying that all the  strictly positive, and thus that 
%the one irregular vertex is internal to the patch. 
%When the inequality on $s_i$ are met in the equality sense, the irregular vertex appears on the boundary, which can be acceptable or not . 

\section{Resulting quadrangulation}

In addition to stating the conditions that must be met for a patch to be CC-able (either strictly or non-strictly), our construction also provides a closed-form description of the %possible 
resulting quadrangulation.  

%\subsection{Closed-form construction}

%For the cases $n=2,3,4,5,6$ and $7$,

%in the form of the values of $s_i$ are fully determined.
%This is not the case for $n=8$, which admit multiple solutions (if any);
%as a generalization, this is also the case for any $n = 2^k$, for any natural $k$.

The set of values $s_i$ can be understood as a compact and complete way to describe the internal tessellation
of the patch.
Specifically, they describe the grid-sizes of the regular rectangular areas constructed on the corners of the original patch; or, equivalently, the number of edges of the sub-sides splitting each side, and, thus, the topological position of the singularity inside $P$
(see Figure ~\ref{fig:schema}).

The Equations (\ref{eq:twoSi},\ref{eq:triSi},\ref{eq:pentaSi},\ref{eq:hexaSi},\ref{eq:eptaSi},\ref{eq:octaSi}) define the values of $s_i$ as a closed function of $e_i$ for the cases of $n = 2,3,5,6,7,8$ respectively. In the last case, the equations use two arbitrary values $k_i$, 
%which reflect the choice among all possible quadrangulation, and 
which must be chosen inside the intervals defined in closed-form by Eq. (\ref{eq:octaKi}), spanning all and only the possible solutions.

\section{Discussion}

In this work, we identified a problem statement for a self-contained simple task, useful in the context of Geometry Processing, and derived a complete answer. Specifically, we expressed in closed form the conditions for CC-ability,  the description of the resulting quadrangulation, and the set of available choices (if multiple solutions are possible)

\paragraph*{About the uniqueness of the solution}
As we have shown, the sought single-irregular-vertex quadrangulation can be non-unique for $n=8$. 
This contradicts the commonly held notion that, in a quadrangulation, no irregular vertex can be ``moved alone'' (e.g. \cite{editMesh2011}), meaning that its topological position  inside a region cannot be modified without also affecting either the tessellation of the region boundary or the topological position of another irregular vertex in the same region. As it turns out, valency-8 vertices are exceptional in that they \emph{can} be ``moved alone'', 
as exemplified in Figure~\ref{fig:example8}. As we have shown, this is the lowest valency for which this happens (but it also occurs for valencies 12, 16, 20, and so on).
%generalizes to irregular vertices with a valency that is a multiple of 4.

\paragraph*{Usability} 
The property of being CC-able can be easily embedded in optimization systems,  because it is expressed in closed form as a set of linear inequality, equality, or parity   constraints, which can be enforced for example in 
%Linear Programming, Integer Linear Programming, 
convex numerical solvers.

\paragraph*{Potential applications}
While the main motivation for our analysis stems from Coarse-layout construction (see discussion in Section \ref{sec:motivations}),
another potential context is that of shape modelling, where digital modellers often define 3D shape high-level description, leaving an automatic system in the background to deal with the minutiae of the surface tessellation (for example \cite{Takayama2014}), such as in the software suits 3D Coat (PilgWay), Z-Brush (Pixologic) and others. Due to the mentioned direct relationship with Catmull-Clark subdivision, our formulation can be a useful tool in the context of ``reverse subdivision'' (for example, \cite{subdiv}), where the objective is to seek a subdivision surface approximating a given shape.

More broadly speaking,
the generality of the analyzed problem statement leads us to believe that our closed-form formulation can find numerous applications. %in numerous Geometry Processing or Computational Geometry problems.

%\begin{equation}
%
%\left( \begin{matrix} 
% 0 & 1 & 0 & 0 & 0 & 0 & 0 & 1 \\
% 1 & 0 & 1 & 0 & 0 & 0 & 0 & 0 \\ 
% 0 & 1 & 0 & 1 & 0 & 0 & 0 & 0 \\ 
% 0 & 0 & 1 & 0 & 1 & 0 & 0 & 0 \\ 
% 0 & 0 & 0 & 1 & 0 & 1 & 0 & 0 \\
% 0 & 0 & 0 & 0 & 1 & 0 & 1 & 0 \\
% 0 & 0 & 0 & 0 & 0 & 1 & 0 & 1 \\ 
% 1 & 0 & 0 & 0 & 0 & 0 & 1 & 0 \\
%\end{matrix} \right)
% %
%\left( \begin{matrix} 
% s_0 \\
% s_1 \\ 
% s_2 \\ 
% s_3 \\ 
% s_4 \\
% s_5 \\ 
% s_6 \\ 
% s_7 
%\end{matrix} \right)
% =
%\left( \begin{matrix} 
% e_0 \\
% e_1 \\ 
% e_2 \\ 
% e_3 \\ 
% e_4 \\ 
% e_5 \\ 
% e_6 \\ 
% e_7 
%\end{matrix} \right)
%
% \end{equation}
%\bibliographystyle{ieee_fullname}

\bibliographystyle{acm}

\bibliography{main}

\end{document}